\documentclass[10pt,conference,a4paper]{IEEEtran}
\usepackage{times,amsmath,epsfig}
\usepackage[english]{babel}
\usepackage[latin1]{inputenc}
\usepackage[T1]{fontenc}
\usepackage[latin1]{inputenc}
\usepackage{float}
\usepackage{algorithmic}
\usepackage{amssymb}
\usepackage{cite}
\input{Qcircuit}

\newtheorem{theorem}{Teorema}[section]
\newtheorem{definition}[theorem]{Definition}

\title{Quantum Communication Complexity of\\ Quantum Authentication Protocols}
\author{
{Elloá B. Guedes and Francisco M. de Assis}
\vspace{1.6mm}\\
\fontsize{10}{10}\selectfont\itshape
IQuanta -- Institute for Studies in Quantum Computation and Quantum Information\\
Federal University of Campina Grande\\
Rua Aprígio Veloso, 882 -- Campina Grande -- Paraíba -- Brazil\\
\fontsize{9}{9}\selectfont\ttfamily\upshape
elloaguedes@gmail.com, fmarcos@dee.ufcg.edu.br\\
}

\begin{document}
\selectlanguage{english}
\maketitle

\begin{abstract}
In order to perform Quantum Cryptography procedures it is often essencial to ensure that the parties of the communication are authentic. Such task is accomplished by quantum authentication protocols which are distributed algorithms based on the intrinsic properties of Quantum Mechanics. The choice of an authentication protocol must consider that quantum states are very delicate and that the channel is subject to eavesdropping. However, even in face of the various existing definitions of quantum authentication protocols in the literature, little is known about them in this perspective, and this lack of knowledge may unfavor comparisons and wise choices. In the attempt to overcome this limitation, in the present work we aim at showing an approach to evaluate quantum authentication protocols based on the determination of their quantum communication complexity. Based on our investigation, no similar methods to analyze quantum authentication protocols were found in the literature. Pursuing this further, our approach has advantages that need to be highlighted: it characterizes a systematic procedure to evaluate quantum authentication protocols; its evaluation is intuitive, based only on the protocol execution; the resulting measure is a concise notation of what resources a quantum authentication protocol demands and how many communications are performed; it allows comparisons between protocols; it makes possible to analyze the communication effort when an eavesdropping occurs; and, lastly, it is likely to be applied in almost any quantum authentication protocol. To illustrate the proposed approach, we bring results about its application in ten existing quantum authentication protocols ( data origin and identity authentication). Such evaluations increase the knowledge about the existing protocols, presenting their advantages, limitations and contrasts.
\end{abstract}

\section{Introduction} \label{sec:intro}

Quantum Cryptography comprehends both Quantum Mechanics and Information Theory. As the classical Cryptography, its objectives goes beyond confidentiality and include methods to provide data integrity, non repudiation and authentication \cite{Delfs:IntroToCryptography}. In particular, \emph{authentication} concerns the procedures to verify the origin of some data or to verify the identity of a party in the communication. For this reason, authentication is subdivided in \emph{data origin authentication} and in \emph{entity authentication}.

In the Quantum Cryptographic domain, authentication is performed by \emph{quantum authentication protocols} which are \emph{distributed algorithms} based on the intrinsic properties of Quantum Mechanics. Quantum authentication protocols differ from the classical ones in at least three aspects: they are not based in some computational difficulty; they don't allow information copy; and, they may allow eavesdropping detection \cite{Li:JournalAuthenticationEntangled}. In association with quantum key distribution protocols, they play an important role in providing trusty quantum communication in the presence of eavesdroppers.

Given the importance of authentication in quantum communication, several protocols for quantum authentication have been proposed in the literature \cite{Barnum:AuthenticationPioneiro,Yang:MessageAuthenticationNP,Curty:AuthentiationClassicalMessages,Li:MessageAuthentication,Kanamori:AuthenticationSuperposition,Zeng:QAuthenticationProtocol,Li:JournalAuthenticationEntangled,Zhang:QuantumAuthenticationEntangled,Barnum:EntaglementCatalysis,ZengZhang:IdentityVerification}. They use different procedures and resources of the Quantum Mechanics to perform the authentication, such as: EPR pairs, superposition, catalysis, unitary operations, among others.

Considering that quantum states are very delicate \cite{Gottesman:QuantumStatesDelicate}, a crucial concern in the adoption of a certain quantum authentication protocol is the number of communications performed. A protocol that minimizes such number but that still ensures secure authentication can be considered well suited to practical scenarios of Quantum Cryptography. However, little is known about the existing protocols in this perspective and this lack of knowledge can unfavor comparisons and wise choices of quantum authentication protocols.

Taking into account these concerns, in this paper we propose an approach of evaluation and classification of quantum authentication protocols based on the determination of their \emph{Quantum Communication Complexity} -- a measurement of the amount of communications carried out between the parties in order to accomplish some distributed task \cite{Wolf:Survey:QuantumCommunicationAndComplexity}. As far as we know, no similar approaches to analyze quantum authentication protocols were found in the literature.

The main result of our proposition is that it turns out to be possible the determination of the resources required by a certain protocol and the realization of systematic comparisons between different quantum authentication protocols. It is important in practical scenarios of Quantum Cryptography because authentication protocols can be ranked and chosen according to the communication resources available.

Furthermore, this paper presents the application of our proposed methodology in ten quantum authentication protocols. This illustrates how the approach can be applied and brings new results about the characteristics and advantages of quantum authentication protocols existing in the literature.

The rest of the paper is organized as follows. Section \ref{sec:quantumComComplexity} introduces the basics concepts of the Quantum Communication Complexity. Section \ref{sec:quantumAuthentication} presents the formalism and concepts regarding quantum authentication protocols. Section \ref{sec:approach} introduces the approach proposed and Section \ref{sec:evaluation} shows the results of our analysis in some existing quantum authentication protocols. Lastly, Section \ref{sec:remarks} draws the conclusions and suggestions for future work.

\section{Quantum Communication Complexity} \label{sec:quantumComComplexity}
With the advents of telegraph and telephone in the mid-twentieth century, there was an urge to perform the tasks of store, transmit and process data. Motivated by these practical problems, the \emph{Information Theory} was proposed \cite{Cover:ElementsInformationTheory}. Shannon laid its foundations with an article entitled ``\emph{A Mathematical Theory of Communications}'' \cite{Shannon:OriginalPaper} where he defined precisely what is information and how to measure it, and also demonstrated the existence of codes to error-free communication when the channel capacity is respected.

It is important to emphasize that the necessity to communicate arises when two or more parties need to collaborate jointly to accomplish a task that none of them can perform alone. In Information Theory, the objective is to study \emph{how} this communication must be performed -- which codes to use, how to deal with noisy channels, and so on. Another perspective that can be taken into account when a communication needs to be carried out is \emph{what} needs to be communicated. This is the object of study of Communication Complexity \cite{Yao:ComplexidadeOriginal,Kushilevitz:LivroCommunicationComplexity}.

With the growing adoption of quantum channels in communications, Yao \cite{Yao:QuantumCircuitComplexity} was a pioneer in considering the concerns of Communication Complexity in this domain. It was needed to understand the implications of a communication that makes use of Quantum Mechanics resources with the purpose to answer a central question: ``Are there any advantages in the quantum communication model when compared to the classical existing ones?''.

In the study of Quantum Communication Complexity it is considered that two \emph{parties}, say Alice and Bob, are interested in evaluate a certain $f(x,y)$, where $x$ is known only by Alice and $y$ is known only by Bob. Alice and Bob must exchange information through a supposed error-free quantum \emph{channel} according to some \emph{protocol} which can be understood as a distributed algorithm. The main interest in this scenario is the \emph{amount of communication} necessary to the parties accomplish their task.

According to the resources of Quantum Mechanics available to Alice and Bob, there are three variants of the Quantum Communication Complexity model that can be considered:

\begin{enumerate}
\item \textbf{Yao's model} \cite{Yao:QuantumCircuitComplexity,Kremer:Dissertacao}. This model considers a \emph{quantum channel} who will enable exchanges of qubits between Alice and Bob. Each party of the communication interacts with the channel via unitary operations, depositing qubits that can be accessed by the other part also via unitary operations with the channel. When Alice and Bob determine precisely the value of $f(x,y)$ the protocol ends and the number of communications is considered. In this variant, the Quantum Communication Complexity of a function $f$ is denoted by $Q(f)$;
\item \textbf{Cleve and Buhrman's model} \cite{Cleve:ModeloEmaranhamento}. This model considers the existence of \emph{prior entanglement} between the parties and allows the exchange of information via a classical channel. In this variant, the number of entangled pairs is not considered, just the number of classical bits exchanged. In this variant, the quantum communication complexity of a function $f$ is denoted by $C^{\ast}(f)$. It is important to emphasize that this model is well suited to analyze protocols where superdense coding and teleportation are used;
\item \textbf{Hybrid Model}. This variant combines the characteristics of the previous two: there are \emph{entangled pairs} available to the parties of the communication, a \emph{quantum channel}, and also a \emph{classical channel}. In the determination of the Quantum Communication Complexity the entangled pairs used are not considered, just the further information exchanged. In this variant, the Quantum Communication Complexity of a function $f$ is denoted by $Q^{\ast}(f)$.
\end{enumerate}

In all the three variants considered, we assume that the parties must determine precisely the value of $f(x,y)$, i.e., the function $f$ must be evaluated with probability of error equal to zero. A more detailed description of these models and alternative definitions that enables a limited error can be found in the surveys of Wolf \cite{Wolf:Survey:QuantumCommunicationAndComplexity} and Brassard \cite{Brassard:SurveyPhysics}.

The theory of Quantum Communication Complexity has still open questions that need to be enlighten, such as about the existence or not of an exponencial gap between Classical and Quantum Communication Complexity for all functions. Despite this, applications of Quantum Communication Complexity are growing every day. Results on quantum formula \cite{Yao:QuantumCircuitComplexity},  finite automata size \cite{Klauck:AutomataSize}, data structures \cite{Ventakesh:CommunicationsDataStructure}, and on security of quantum key distribution \cite{BenOr:QKDProof} have already been developed.

The most interesting aspect of Quantum Communication Complexity is that the advantage provided by Quantum Mechanics has been established rigorously. This is in sharp contrast with the field of Quantum Computing, in which it is merely believed that Quantum Mechanics allows for an exponential speedup in some computational tasks \cite{Brassard:SurveyPhysics}.

\section{Quantum Authentication Protocols} \label{sec:quantumAuthentication}
Authentication is a well-studied area of classical cryptography. It is concerned with assuring that a communication is authentic. In the case of a single message, the function of the authentication is to assure the recipient that the message is from the party that it claims to be from. In the case of an ongoing interaction, two aspects are involved: first, at an initial time, the objective is to assure that the two parties are authentic, that is, that each is the party that it claims to be; after that it must be assured that the connection is not interfered in such a way that an eavesdropper can masquerade as one of the two legitimate parties for the purposes of unauthorized transmission or reception \cite{Stallings:LivroCryptography}.

Aiming to provide these two functions, authentication must be achieved in two branches:

\begin{enumerate}
\item \textbf{Data Origin Authentication}. Enables the recipient to verify that the message has not been tampered \emph{en route} and that it originate from the expected sender;
\item \textbf{Identity Authentication}. Enables the recipient to verify that a sender is who his claims to be. If some security conditions are guaranteed, it also enables the recipient to ensure that no one else is impersonating the true sender.
\end{enumerate}

\begin{figure*}[t!]
  \includegraphics[width=\textwidth]{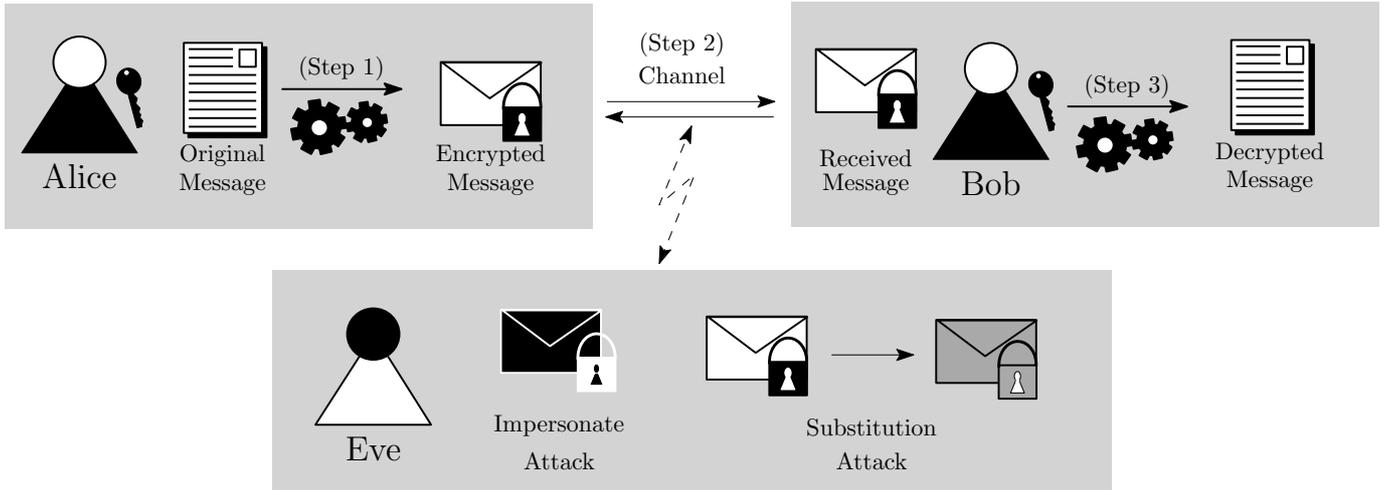}\\
  \caption{Symmetric Key Model of Authentication.}\label{fig:modelAuthentication}
\end{figure*}

To illustrate how authentication works consider a simple symmetric key model of authentication consisted of a sender (Alice), a receiver (Bob), and an eavesdropper (Eve) as illustrated in the Figure \ref{fig:modelAuthentication}. The objective in this model is to enable Bob to authenticate a message sent by Alice. To do so, in a previous moment Alice and Bob securely share a key $k$ that will be used to authentication.

Alice encrypts the original message $m$ with the key $k$ using an algorithm $E$, producing $m_c = E(m,k)$ (Step $1$). Alice sends $m_c$ to Bob through an insecure channel which is being eavesdropped by Eve (Step $2$). It is assumed that Eve can observe all the information transmitted from the sender to the receiver and also that, in general, she knows even the original message, but not the key used to encrypt it.

There are two kinds of possible attacks by the opponent: the \emph{impersonation attack} in which Eve sends a message in the hope that it will be accepted by the receiver Bob as a valid one; and the \emph{substitution attack} in which he opponent observes a transmitted message and then replaces it with another message.

Upon receiving a message, Bob will use his key $k$ and a decryption algorithm $D$ to try recover the original message, i.e., he will perform $D(m_c,k)$ (Step $3$). If no tampering occured, Bob will obtain a pair $\left\langle m, 1\right\rangle$ where $m$ is the original message sent by Alice and $1$ indicates that the authentication was successful. Otherwise, Bob must discard and ask Alice to send again  \cite{Delfs:IntroToCryptography,Tilborg:Enciclopedia}.

In the quantum setting, despite the same idea, authentication is performed sightly differently. Information is now \emph{physical} and for that reason the laws of the Quantum Mechanics must be obeyed. In this domain, authentication is thus performed by \emph{quantum authentication protocols} which define how to represent information and what to be sent by each party.

To send a message $m$ to Bob according to a quantum authentication protocol, Alice encodes $m$ with a certain code before sending it. However, if the same code is always used, Eve can simply create errors that the code cannot detect, so they must use one of a family of codes which detect different kinds of errors. The key $k$ now tells them which code to use. Since Eve doesn't know $k$, she doesn't know which errors the code detects, and no matter what she tries to do, she has a good chance of getting caught. Besides, Alice and Bob must also encrypt the quantum state in order to avoid undetectable changes in the quantum state performed by Eve \cite{Barnum:AuthenticationPioneiro}.

In comparison with the classical protocols, the quantum ones differ in several potentially useful ways. The primary contrast between them regards how information is represented and exchanged -- while the classical protocols are restricted to bits, in the quantum scenario the parties can use qubits, entangled particles and even bits. Another difference regards the action of the eavesdropper. In the classical scenario, Eve would break into the classical storage area and copy Bob's key without leave any evidence. After that, she would use it to communicate with Alice who might now realize anything wrong for quite a while. However, in the quantum scenario it is improbable to occur. The no-cloning theorem not only makes undetected theft of key more difficult, but also protects a stolen key from dissemination \cite{Barnum:EntaglementCatalysis}. Furthermore, in the quantum domain, authentication implies encryption which is not always true in the classical scenario \cite{Barnum:AuthenticationPioneiro}.

Quantum authentication plays a major role in Quantum Cryptography because they precede the execution of the so called quantum key distribution (QKD) protocols. Such protocols provide the conditions to the parties produce a shared random secret key known only to them, which can then be used to encrypt and decrypt messages \cite{Chuang:Biblia}. Since QKD protocols require previous authentication and considering the recent successful results about their implementation over long distances \cite{Hughes:Implementation}, there was a motivation for the presentation of a formal definition \cite{Barnum:AuthenticationPioneiro} and for the proposition of quantum authentication protocols in the literature \cite{Yang:MessageAuthenticationNP,Curty:AuthentiationClassicalMessages,Li:MessageAuthentication,Kanamori:AuthenticationSuperposition,Zeng:QAuthenticationProtocol,Li:JournalAuthenticationEntangled,Zhang:QuantumAuthenticationEntangled,Barnum:EntaglementCatalysis,ZengZhang:IdentityVerification}.

\section{The Proposed Approach} \label{sec:approach}
As stated in the Section \ref{sec:quantumAuthentication}, the purpose of the interaction of the two parties in a quantum authentication protocol is to enable the verifier Bob to evaluate $D(m_c,k) = \left\langle m, \textrm{valid} \right\rangle$, where $m$ is the original message sent by Alice. Introducing some formalism, a quantum authentication protocol can be represent as a function $f_A : M \times K \rightarrow \left\{ 0 , 1\right\}$ that evaluates to $1$ if $D(m_c,k) = \left\langle m, \textrm{valid} \right\rangle$, and to $0$ otherwise. To make a distinction between data origin and identity authentication protocols, we will denote the corresponding functions by $f_D$  and $f_I$, respectively.

Different quantum authentication protocols can implement $f_D$ and $f_I$ \cite{Barnum:AuthenticationPioneiro,Yang:MessageAuthenticationNP,Curty:AuthentiationClassicalMessages,Li:MessageAuthentication,Kanamori:AuthenticationSuperposition,Zeng:QAuthenticationProtocol,Li:JournalAuthenticationEntangled,Zhang:QuantumAuthenticationEntangled,Barnum:EntaglementCatalysis,ZengZhang:IdentityVerification}.
Therefore, it is natural to look for ways to compare them in order to choose one that is better or more adequate to certain available resources. Despite the differences between the implementations of all these protocols, they share a common characteristic -- communications are performed between the parties.

Taking the communications between the parties into account and remembering that quantum states are delicate \cite{Gottesman:QuantumStatesDelicate} and also that the channel is subject to eavesdropping (that may affect the quantum states), a quantum authentication protocol can be considered good if it \emph{minimizes the number of communications} between the parties but still assures a \emph{secure authentication}. In order to meet the stated criteria, we present an approach based on the analysis of the quantum communication complexity of a quantum authentication protocol, defined as follows:

\begin{definition} (\textbf{Quantum Communication Complexity of a Quantum Authentication Protocol}) Let $\mathcal{P}$ be a quantum authentication protocol between two parties $A$ and $B$ that computes a function $f_A : M \times K \rightarrow \left\{ 0,1 \right\}$ (that can be $f_D$ or $f_I$). The Quantum Communication Complexity of $f_A$ under $\mathcal{P}$ is the minimum number of communications between $A$ and $B$ to ($i$) allow both parties to compute $f_A(m_c,k)$ where $m_c$ and $k$ are the worst case over all inputs $M \times K$ (i.e., the cost of the protocol is maximal), and to ($ii$) avoid an eavesdropper Eve to create any $m'$ such that $f_A(m',k) = 1$ or to evaluate $f_A(m_c,k) = 1$.
\end{definition}

Regarding the assumption that the computational power is unlimited, the security that avoids Eve to create any $m'$ such that $f_A(m',k) = 1$ or to evaluate $f_A(m_c,k) = 1$ is desired to be unconditional. An exponentially low probability of success in the attacks is also acceptable, but there are quantum authentication protocols that still rely on unproven computational difficulties. Protocols with this last mentioned characteristic have weaker security when compared to the two others.

It is important to emphasize that the definition of quantum communication complexity of a quantum authentication protocol used by our approach is sightly different from the original one of quantum communication complexity. In the original definition if one of the parties is able to determine the evaluation of the function, it is enough to communicate it to the other party. However, in our definition it is not allowed: the presence of the eavesdropper avoids Alice to send $m$ directly to Bob or to perform any communication that may reveal $k$.

Another consideration that must be made regards the resources of Quantum Mechanics required by each protocol. In the determination of the quantum communication complexity of a quantum authentication protocol, each protocol will be evaluated according to one of the three variants -- Yao, Cleve and Buhrman, or Hybrid -- presented in the Section \ref{sec:quantumComComplexity}. Moreover, for reference, the quantum communication complexity will be denoted in function of the size of the key and of the message.

Our approach, therefore, can be described in a straightforward way considering the definitions previously presented in the Sections \ref{sec:quantumComComplexity} and \ref{sec:quantumAuthentication}. To evaluate a quantum authentication protocol firstly it is necessary to determine the Quantum Mechanics resources required -- quantum bits, classical bits and/or previously shared entangled pairs. After that, with a key of $n$ bits and a message of $m$ bits (if they exist), perform an execution of the protocol supposing that no impersonation nor substitution attacks occur. From this execution, it is necessary to verify how much information was exchanged between the parties to compute $f_A$. The information exchanged must be expressed in terms of $n$ and $m$. As the final step, according to the resources identified in the initial moment, the information exchanged must be represented in accordance with the respective model, i.e., using $Q$, $C^{\ast}$ or $Q^{\ast}$.

Upon analyzing a quantum authentication protocol according to our approach, the final result is a concise notation of what resources it demands and how many communications are performed, considering a key and a message of a generic size. If the exact number of communications may vary depending on certain situations, this approach also allows an asymptotic notation where the upper and lower bounds as well as best, average, and worst cases can be individually analyzed.

The asymptotic notation can also be used to measure the effort necessary so the parties can recover from an eavesdropper attack. If the attack is detectable, it is necessary to measure the resources to recover from it and to still ensure the authentication. A certain quantum authentication protocol may require the parties to discard and restart from the initial point while other can reuse the non-tampered data, saving on the number of communications. If the evaluation of a protocol will be made considering the action of an eavesdropper, its quantum communication complexity will be denoted as $Q_E$, $C^{\ast}_{E}$, or $Q^{\ast}_{E}$ according to the most adequate model.

The comparisons between quantum authentication protocols are also possible thanks to the proposed approach. Following the procedures previously described, the evaluation of the quantum communication complexity of each protocol must be performed independently. After that, the results obtained must be grouped according to the respective model and, then, ordered. The lower result in each group indicates the protocol that requires less communications to provide authentication, i.e., that best fits the previously state criteria. It is also important to emphasize that no comparisons between protocols classified under different groups are possible -- the resources involved are of different nature and their comparison can lead to misleading conclusions.

Based on our investigation in the literature, no similar methods to analyze quantum authentication protocols were found. So, the approach presented contributes to overcome this limitation. Pursuing this further, there are advantages of our approach that need to be highlighted: it characterizes a systematic procedure to evaluate quantum authentication protocols; its evaluation is intuitive, based only on the protocol execution; the resulting measure is a concise notation of what resources a quantum authentication protocol demands and how many communications are performed; it allows comparisons between protocols; it makes possible to analyze the communication effort when an eavesdropping occurs; and, lastly, it is likely to be applied in almost any quantum authentication protocol.

\section{Evaluating the Quantum Communication Complexity of Quantum Authentication Protocols} \label{sec:evaluation}
In order to illustrate the proposed approach, this section shows the evaluation of the quantum communication complexity of quantum authentication protocols existing in the literature. The results obtained and the conclusions achieved are presented in the following subsections according to the purpose of the protocol -- data origin authentication or identity authentication.

\subsection{Results for Data Origin Authentication}

The first proposal of a quantum authentication protocol was made by Barnum et al. \cite{Barnum:AuthenticationPioneiro}. In their protocol, Alice and Bob use purity testing codes and make a prior agreement on some parameters that will be used (two keys, purity code, and syndrome). After that, they carry out the protocol that enables Bob to authenticate the message sent by Alice in a single communication to him. Considering that this protocol requires only qubits exchanges between the parties, in our approach its analysis will be made with the Yao's model. If a message has $m$ qubits, Alice will send Bob a quantum coded message of $m + n$ qubits, where $n$ is the size of a security parameter as a key. Therefore, the quantum communication complexity of this protocol is $Q(f_D) = m + n$.

In the quantum data origin authentication protocol proposed by Yang et al. \cite{Yang:MessageAuthenticationNP}, Alice wants to send a message composed by a sequence of pure states. To do so, she encodes it with a Goppa code using parameters previously securely shared with Bob. The encoded message that will go through the channel has twice the qubits of the original one. When Bob receives such message he decodes it with an unitary operator built from the parameters of the Goppa code in use. The decoding procedure uses up to five registers where, in particular, the fifth (stores the original message) and the third (stores an authentication flag) must be measured. It should be noticed that this protocol only requires qubits exchanges which implies the analysis of its Quantum Complexity Communication with the Yao's model. Each message with $m$ qubits is codified with a Goppa code using $2 \cdot m$ qubits, therefore, $Q(f_D) = 2 \cdot m$. In conclusion, the security of this protocol relies on the computational hardness of building the decoding operator without the knowledge of the security parameters.

The quantum authentical protocol proposed by Curty and Santos \cite{Curty:AuthentiationClassicalMessages} uses a code to protect the message that will be sent through the channel. Alice and Bob start the protocol already sharing a singlet state. When Alice wants to send a bit of the message to Bob she prepares two qubits in the state $\ket{\phi_i}$, where $\ket{\phi_0}$ and $\ket{\phi_1}$ are orthogonal states and represent the classical bits ``0'' and ``1'', respectively. After that Alice encodes the qubit along with her part on the singlet and send it to Bob. Upon receiving the message, Bob uses a decoding operator and is able to detect, with high probability, when a tampering occurred and when the message is authentic from Alice. In this protocol, the parties use qubits and a previously entangled pair, but no classical communication is performed between them. This implies that the Hybrid model must be used in the analysis. As explained before, to each bit of the classical message that Alice wants to send Bob, she must use a two qubits state. Hence, the quantum communication complexity of this protocol is $Q^{\ast}(f_D) = 2 \cdot m$.

Li and Zhang \cite{Li:MessageAuthentication} present a message authentication protocol that uses previously shared EPR pairs between the parties. When Alice wants to send a bit to Bob she codifies it in a quantum system before sending to Bob. She uses a redundant coding, in which two qubits (in Bell states) codes one bit of information, and performs a CNOT operation with her half of the EPR pair. When Bob receives such state, he also performs a CNOT operation (on his half and on the received qubit) followed by a measurement. Bob is able to recover the original message sent by Alice and also to detect eavesdropping. In the described protocol the parties use entangled pairs and exchange qubits, but they don't perform classical communication. It is possible to conclude, thus, that the model to analyze it is the Hybrid one. Moreover, Alice uses a codification where $1$ bit is codified in $2$ qubits. This way, the quantum communication complexity of this protocol is $Q^{\ast}(f_D) = 2 \cdot m$.

In the four protocols analyzed, two of them were evaluated under the Yao's variant and the other two under the Hybrid variant. Regarding the first group, the Barnum et al. \cite{Barnum:AuthenticationPioneiro} protocol may have smaller quantum communication complexity than the protocol of Yang et al. \cite{Yang:MessageAuthenticationNP} when $n < m$, but the security issues must be considered. The other two protocols have equivalent quantum communication complexity. It is also important to notice that none of the protocols analyzed make use of classical communications.

\subsection{Results for Identity Authentication}

In the identity authentication quantum protocol proposed by Kanamori et al. \cite{Kanamori:AuthenticationSuperposition,Kanamori:Globecom} Alice and Bob share a prior key $K = \left\{ \theta_i : 0 \leq \theta_i < \pi, i = 1, 2, \ldots, n \right\}$ composed of a sequence of angles.  Alice generates a random $n$-bit string $R_A$ and encodes it into a system $\ket{\psi_{R_{A}}}$ of $n$ qubits in orthogonal states ($\ket{0}$ or $\ket{1}$, for instance). In sequence, Alice rotates each qubit $\ket{\psi_{A,i}}$ according to an angle $\theta_i \in K$, encrypting the original quantum state. After that, she sends the resulting state to Bob. Since Bob knows $K$, he decrypts the received state, performs a measurement, and, therefore, recovers $R_A$. Next, Bob generates a random $n$-bit number $R_B$ and a session key $K_S = \left\{ \theta_i' : 0 \leq \theta_i' < \pi, i = 1,2,\ldots, n \right\}$. In sequence, he encodes $R_B$ in a quantum system $\ket{\psi_{R_B}}$ and encrypts it with $K$ and $K_S$ before send it to Alice. Alice decrypts the state with $K$ and performs an exclusive-OR (XOR) operation with the resulting state and $R_A$. She sends the resulting state to Bob who decrypts with $K_S$ and obtains a superposition $\ket{\psi_{R_A \oplus R_B}}$. The last step performed by Bob is a XOR with $R_A$. If he retrieves $R_B$, he can successfully authenticate Alice's identity.

As it can be seen, the protocol proposed by Kanamori et al. just uses a quantum channel and no prior entanglement. In our approach, thus, it will be used the Yao's model to analyze it. Considering that the key $K$ shared between the parties has $n$ bits, and $3$ quantum states of dimension $n$ are exchanged in this protocol, the resulting complexity is $Q(f_I) = 3\cdot n$. One disadvantage of this protocol is that if tampering occurs, a complete repetition of the procedures must be carried out. Despite of that, it generates a session key that can be used later by the parties of the communication.

Zeng and Guo \cite{Zeng:QAuthenticationProtocol} present an identity authentication quantum protocol that is based on symmetric cryptography with EPR pairs previously shared between the parties. In their protocol, Alice and Bob share a prior key $K_1$ of $n$ bits. From the key, they derive a serie of measurements $M_{K_1}$ in a rectilinear or diagonal basis. Initially, Alice performs a serie of measurements in her half of the EPR par with $M_{K_1}$. In his turn, Bob measures his half with $M_{K_1}$ and with a random series of measurements $M$. If eavesdropping occurred, Alice and Bob can detect by showing each other certain results of their measurements in common. Bob and Alice therefore change their results via classical symmetric key cryptography and can authenticate the identity of each other.


The described protocol is able to authenticate both the identities of Alice and Bob. To do so, it uses EPR pairs and classical cryptography. Regarding this last point, in particular, unconditional security cannot be guaranteed. Taking into account the resources required by this protocol, its quantum communication complexity must be evaluated with Cleve and Buhrman's variant. The communications performed between Alice and Bob are encrypted versions of their measurement results who require $2 \cdot n  + s$ bits of classical information, where $s$ is a security parameter. Thus, according to our approach, the quantum communication complexity of this protocol is $C^{\ast}(f_I) = \Omega(2 \cdot n)$.

Another identity authentication quantum protocol was proposed by Li and Barnum \cite{Li:JournalAuthenticationEntangled}. This protocol uses EPR pairs between the parties as the identification token. In this protocol, Alice and Bob previously share $n$ EPR pairs and create an EPR pair associated to each of them. These auxiliary pairs will be measured in the  Bell basis at the end of the process. If the Alice party is legitimate and no tampering occurred, Bob will get one of the Bell states previously expected.

One interesting aspect of the protocol of Li and Barnum is that no previous key is shared between the parties, just entangled qubits. It should also be noticed that no classical communication is required although qubits exchanges occurs. Therefore, this protocol must be analyzed according the Hybrid model. The resulting quantum communication complexity  is related with the numbers of communications required to produce the EPR pairs: $Q^{*}(f_I) = 2 \cdot n$.

Zhang, Li and Guo \cite{Zhang:QuantumAuthenticationEntangled} present a quantum identity authentication protocol that uses previously shared EPR pairs and a quantum channel. In their protocol, Alice acts as an identifier, Bob as a verifier, and they share an angle $\theta$ that will be helpful in the prevention of impersonation. When the protocol starts, Alice and Bob rotate $2 \cdot k$ entangled pairs by $\theta$. After that, Bob creates $k'$ ($k' \leq k$) qubits in an arbitrary pure state, denoted by $\ket{\psi_i}$, and send it to Alice who will perform CNOT operations controlled by her half of the entangled pair. Alice sends that particles back to Bob who uses his corresponding particles of the entangled pair to do a CNOT operation, making $\ket{\psi_i}$ turn back to the original state. Bob then performs a measurement in the basis $\{ \ket{\psi_i}, \ket{\psi_i}^\perp \}$ and checks if the results obtained are in accordance with what is expected. If the measurements passes the test, Bob can authenticate Alice.

In this protocol, the operations performed are the strength against eavesdroppers. Besides, the EPR pairs shared between the parties are intact after the execution and can be reused to help Alice authenticate Bob, for instance. To analyze this protocol, the Hybrid variant will be used since it makes use of previously entangled qubits and of a quantum channel. Considering that the state $\ket{\psi_i}$ has $n$ qubits and that it is sent to Alice and then back to Bob, the quantum communication complexity of this protocol is $Q^{\ast}(f_I) = 2 \cdot n$. It is also important to emphasize that no classical communication is carried out.

Barnum \cite{Barnum:EntaglementCatalysis} proposes a quantum identity authentication protocol that exploits the phenomenon of entanglement-catalyzed transformations between pure states. Alice and Bob share a catalyst state $\ket{\phi}$, and there are incommensurate states $\ket{\phi_1}$ and $\ket{\phi_2}$ such that in the presence of the catalyst, $\ket{\phi_1}$ can be converted to $\ket{\phi_2}$ while retaining $\ket{\phi}$. When Alice wants to authenticate, Bob prepares $\ket{\phi_1}$ and sends half of it to her. They go through the steps, involving local measurements, one-way communication of measurement results, and local operations conditional on those measurements results, which convert $\ket{\phi_1}$ to $\ket{\phi_2}$. This protocol involves qubits exchanges and classical communication and, thus, the model to analyze it is the Hybrid one. However, despite the security of this protocol, the number of communications may vary depending on the steps to transform a certain $\ket{\phi_1}$ into a $\ket{\phi_2}$. For this reason, the quantum communication complexity of this protocol cannot be precisely determined. It just can be said that $Q^{\ast}(f_I) = \Omega(n)$ and that this may not be a tight lower bound.

The protocol proposed by Zeng and Zhang \cite{ZengZhang:IdentityVerification} uses a trusted center to help the legitimate users to authenticate identity. The trusted center sets up a quantum channel between Alice and the center and between Bob and the center. The center generates the same two entangled pairs to Alice and Bob, keeping half of each. Similarly to BB84, Alice and Bob measure their particles with a randomly chosen basis (horizontal-vertical or diagonally polarized) and share the basis used for the measurements, creating a session key -- so, in this protocol, both authentication and QKD are implemented. The resources used are previously entangled pairs, quantum and classical communication what implies in the analysis according to the Hybrid model. The number of communications cannot be determined precisely because it depends on the size of the key. Apart from it, a lower bound of $Q^{\ast}(f_I) = \Omega(4n)$ can be determined. It is important to mention that this protocol is provably secure.

Once the evaluation of the quantum communication complexity of each quantum identity authentication protocol was performed, it is possible to draw some conclusions about them. A common characteristic of all of these protocols is that the number of communications performed is a polynomial in the size of the key. Regarding the models considered, just the protocols from Li and Barnum \cite{Li:JournalAuthenticationEntangled},  Zhang et al. \cite{Zhang:QuantumAuthenticationEntangled}, Barnum \cite{Barnum:EntaglementCatalysis} and Zeng and Zhang \cite{ZengZhang:IdentityVerification} fall in the same variant. The protocols of Li and Barnum \cite{Li:JournalAuthenticationEntangled} and Zhang et al. \cite{Zhang:QuantumAuthenticationEntangled} have equivalent quantum communication complexity. Considering the Barnum's protocol \cite{Barnum:EntaglementCatalysis}, since its quantum communication complexity is highly related to the states used, it is not possible to determine its performance in contrast with the others. But, despite the security, the Zeng and Zhang's protocol \cite{ZengZhang:IdentityVerification} is the one which may perform more communications among the protocols analyzed. Additionally, just the protocol of Kanamori et al. protocol \cite{Kanamori:AuthenticationSuperposition} makes use exclusively of qubits exchanges.

\section{Final Remarks} \label{sec:remarks}
In this paper we presented an approach to evaluate quantum authentication protocols based on the determination of their quantum communication complexity. Our proposal characterizes a systematic procedure to analyze such protocols, allowing comparisons between them and also providing a concise notation of what resources a certain quantum authentication protocol demands.

So far to our knowledge, no similar approaches were found in the literature.  The proposed approach aims to overcome this limitation and also contributes to provide a big picture of the existing quantum authentication protocols. In this context, it helps the identification of the efforts necessary to the proposition of better protocols and may lead new researches in this way.

In the attempt to illustrate the proposed approach, we surveyed the literature and analyzed ten existing quantum authentication protocols. From the quantum data origin authentication protocols, we concluded that two of them have analogous quantum communication complexity according to the Hybrid model and the other two, analyzed under the Yao's model, may distinguish according to the size of the key used in one of them. In the quantum identity authentication protocols covered, it was not possible to determine precisely the quantum communication complexity of all of them, but lower bounds were given in such cases. The resulting analysis helped in the identification of two protocols with equivalent quantum communication complexity and concluded that the Zeng and Zhang's protocol \cite{ZengZhang:IdentityVerification} may require the most communications between all of them. In both categories of quantum authentication protocols investigated, it was possible to conclude that few of them exploits classical communications. The evaluations performed helped in increase the knowledge about the existing literature.

In future works we aim to extend our research to other existing quantum authentication protocols. We also would like to increment the presented analysis, including results about the quantum communication complexity when attacks occur. An open question resultant of this work, in particular, is if it is possible to provide a secure authentication with quantum protocols approximating the Holevo bound, i.e., optimizing the number of communications performed.

\section*{Acknowledgements}
The authors gratefully acknowledge the IQuanta and the financial support rendered by the CNPq \mbox{(Grant \#141446/2011-0)}.

\renewcommand\refname{References}
\bibliographystyle{IEEEtran}
\bibliography{ref}

\end{document}